\documentclass[graybox, envcountchap]{svmult}

\usepackage{mathptmx}        
\usepackage{amsmath}
\usepackage{amssymb}
\usepackage{color}
\usepackage{helvet}          
\usepackage{courier}         
\usepackage{dirtree}

\usepackage{makeidx}        
\usepackage{graphicx}        
\usepackage{subfig}

\usepackage{multicol}        
\usepackage[bottom]{footmisc}

\usepackage{hyperref}        
\hypersetup{colorlinks=true,urlcolor=blue}

\usepackage[misc]{ifsym}

\makeindex             

\begin{document}


\title{High-Resolution X-Ray Spectroscopy of Supernova Remnants}
\author{Satoru Katsuda}
\institute{Satoru Katsuda (\Letter) \at Saitama University, 255 Shimo-Okubo, Sakura, Saitama 338-8570, Japan, \email{katsuda@mail.saitama-u.ac.jp}}
%
%
\maketitle

\abstract{Thermal X-ray spectra from supernova remnants (SNRs) are dominated by a number of line emission from various elements.  Resolving the individual lines is critically important for a variety of scientific topics such as diagnosing high-temperature and low-density non-equilibrium plasmas, identifying spectral features like charge exchange and resonance line scattering, revealing kinematics and elemental abundances of SN ejecta and the circumstellar medium, and studying the interstellar medium or planets' atmospheres from extinction features seen in X-ray spectra of very bright SNRs.  This chapter reviews high-resolution X-ray spectroscopy of SNRs obtained so far.  Most results were derived with dispersive spectrometers aboard Einstein, Chandra, and XMM-Newton satellites.  Because these dispersive spectrometers were slitless, one has to select small objects with angular sizes less than a few arcminutes to successfully perform high-resolution spectroscopy.  Despite this limitation, the three satellites delivered fruitful scientific results in the last few decades.   Arrays of low-temperature microcalorimeters offer excellent opportunities for high-resolution X-ray spectroscopy of SNRs, as they are non-dispersive spectrometers that work for largely extended sources as well as point-like sources.  The microcalorimeter aboard the Hitomi satellite already delivered pioneering results during its short lifetime.  The upcoming X-Ray Imaging and Spectroscopy Mission, which is a recovery mission of Hitomi, will truly open the new discovery window to high-resolution X-ray spectroscopy of SNRs.}


\section{Introduction}
\label{sec:intro}

High-resolution X-ray spectroscopy of various astrophysical objects began with the advent of the Einstein satellite, followed by the Chandra, XMM-Newton, and Hitomi satellites.  The spectrometer aboard Einstein was a curved crystal Bragg spectrometer, named Focal Plane Crystal Spectrometer (FPCS: \cite{Canizares:1977}), with which X-rays of appropriate wavelengths to satisfy the Bragg condition are reflected and detected by one of two redundant imaging proportional counters.  In this case, a crystal produces constructive interference at a single wavelength/angle combination, or narrow range of wavelengths and angles.  Therefore, to take a spectrum over an extended wavelength range, one has to scan through a range of Bragg angles.  On the other hand, Chandra and XMM-Newton carry grating spectrometers.  A great advantage with a grating is that it can diffract radiation in a wide range of wavelengths simultaneously, allowing one to obtain a wide-range high-resolution spectrum at one time.  Chandra carries three sets of transmission gratings: an array of high- and medium-density gratings constitutes the High Energy Transmission Grating Spectrometer (HETGS: \cite{Canizares:2005}), while lower dispersion gratings on the Low Energy Transmission Grating Spectrometer (LETGS: \cite{Brinkman:2000}) provide high spectral resolution out to longer wavelengths.  XMM-Newton carries two identical reflection gratings.  Two separate telescopes have an array of diffraction gratings mounted permanently in the focused beam, constituting the Reflection Grating Spectrometers (RGS: \cite{denHerder:2001}).

All of the diffraction grating spectrometers aboard X-ray astronomy satellites are slitless.  Therefore, these spectrometers work most effectively for point-like sources.  Technically, it is not impossible to make X-ray grating spectrometers that work for diffuse sources as well.  For example, a collimator in front of a telescope can separate a small portion of a spatially extended source out, effectively making it a point-like source.  However, this is practically difficult in X-ray wavelengths for the following reasons.  An aperture that can be limited by a collimator in front of an X-ray telescope is practically no less than $\sim$30 arcminutes, which is too large for a meaningful grating spectroscopy for diffuse sources.  Alternatively, setting a slit behind a telescope may work better.  In fact, such a novel system was recently demonstrated as a sounding rocket instrument, called Marshall Grazing Incidence X-ray Spectrometer \cite{Champey:2022,Savage:2023}, developed to acquire the first ever spatially and spectrally resolved images of solar coronal active region structures in X-rays.  It was designed as a fully grazing-incidence slit spectrograph, consisting of a Wolter-I telescope, slit, spectrometer, CCD camera, and slit-jaw context imager.  In this system, photons passing through the slit are diverging with a range of angles, and thus another grazing incidence telescope with a large aperture populated with a lot of optics covering this entire range of angles is required to refocus the X-rays before incidence on a grating.  Therefore, the spectrometer comprises a matched pair of grazing-incidence parabolic mirrors which refocuses the dispersed light after the slit, and a planar varied-line space grating.  This refocusing process severely limits the effective area, making the system impractical for most X-ray sources.

Although primary targets for the slitless grating spectrometers aboard X-ray astronomy satellites are point-like sources, we should not neglect that they can work for diffuse sources as well, if the angular size of the source is small enough.   For diffuse sources, photons originating at different positions in the sky (along the dispersion direction) strike the grating at different angles.  These photons are detected at wavelength positions shifted with respect to the on-axis source.  Therefore, their measured energies depend not only on the intrinsic photon energy but also on the location in the X-ray source.  This blurs the spectra of extended sources like supernova remnants (SNRs).  However, the XMM-Newton RGS allows one to obtain high-resolution X-ray spectra for somewhat extended sources, thanks to its high dispersion angle which is achieved by the moderate line density reflection grating at grazing incidence.  The wavelength resolution of gratings for a diffuse source is given by $\Delta \lambda \sim d \Delta \theta / m$, where $d$ is the grating period or spacing, $\Delta \theta$ is the angular size of the source in radian, and $m$ is the absolute value of the spectral order.  The effective (projected) grating period of the RGS is $d$sin$\alpha$ $\sim$ $400$ \AA, where $\alpha$ is the angle of incidence.  The projected grating period is 5 and 10 times smaller than that of the Chandra HETG HEG and MEG, respectively.  If the spatial extent of the source is greater than the angular resolution of the X-ray telescope, then the wavelength resolution at a given spectral order depends only on the (projected) grating period (and the spatial extent of the source, of course).  Therefore, the RGS has 5--10 times better resolution than Chandra's gratings for sources with angular extents larger than 30$^{\prime\prime}$.  According to the XMM-Newton Users' Handbook, line broadening ($\Delta \lambda$ in \AA) due to spatial extent ($\Delta \theta$ in arcminutes) is given by $\Delta \lambda = 0.138\,\Delta \theta / m$.  This relation gives spectral resolutions of 3.7\,eV for O He$\alpha$ and 38\,eV at Si He$\alpha$ for $\Delta\theta = 1^\prime$.  Therefore, the spectral resolution for O He$\alpha$ is 20 times higher than that of the non-dispersive CCDs.  More details of the grating spectrometers aboard Chandra and XMM-Newton can be found in \cite{Paerels:2010}.

\begin{table}
\caption{Summary of past high-resolution X-ray spectroscopy of SNRs}
\label{tab:1}       
\begin{tabular}{p{2.5cm}p{2cm}p{1.5cm}p{1cm}p{1.4cm}p{1.4cm}p{2cm}p{1.2cm}}
\hline\noalign{\smallskip}
Name & Distance$^a$ & Age & Type$^b$  & \multicolumn{4}{c}{References for high-resolution X-ray spectroscopy}\\
& (kpc) & (yr) && Einstein & Chandra & XMM-Newton & Hitomi \\
\noalign{\smallskip}\svhline\noalign{\smallskip}
Cygnus Loop & 0.73$\pm0.02$ & 1--2$\times10^4$ & CC  & \cite{Vedder:1986} & --- & \cite{Uchida:2019} & --- \\
RX~J1713.7-3946 & 0.9$\pm$0.6 & 1629 & CC  & --- & --- & \cite{Tateishi:2021} & --- \\
Puppis~A  & 1.3$\pm$0.3 & 4450$\pm750$ & CC & \cite{Winkler:1981a,Winkler:1981b,Canizares:1981} & --- & \cite{Katsuda:2012,Katsuda:2013} & --- \\
SN~1006 & $\sim$2 & 1016 & Ia & --- & --- & \cite{Vink:2003,Vink:2005,Broersen:2013} & --- \\
RCW~86 & 2.2$\pm$0.4 & 1837 & Ia  & --- & --- & \cite{Broersen:2014} & --- \\
Tycho's SNR & 3$\pm$1 & 450 & Ia  & --- & \cite{Millard:2022} & \cite{Decourchelle:2001,Williams:2020} & --- \\
Crab Nebula  & 3.37$^{+4.04}_{-0.11}$ & 968 & CC  & \cite{Schattenburg:1980,Schattenburg:1986} & \cite{Weisskopf:2004,Weisskopf:2011} & \cite{Kaastra:2009} & \cite{Hitomi:2018a} \\
Cas~A & 3.4$^{+0.3}_{-0.1}$  & 342$\pm19$ & CC & \cite{Markert:1983} & \cite{Lazendic:2006,Rutherford:2013} & \cite{Bleeker:2001} & --- \\
G296.1$-$0.5  & 4.3$\pm0.8$  & $\sim$28000 & CC  & --- & --- & \cite{Castro:2011,Tanaka:2022} & --- \\
G21.5$-$0.9 & 4.4$\pm$0.2 & 870$^{+200}_{-150}$ & CC  & --- & --- & --- & \cite{Hitomi:2018b} \\
Kepler's SNR & $\sim$5 & 418 & Ia  & --- & \cite{Millard:2020} & \cite{Blair:2007,Katsuda:2015,Kasuga:2021} & --- \\
G292.0$+$1.8 & 6.2$\pm$0.8 & 3000$\pm60$ & CC  & --- & \cite{Vink:2004} & \cite{Bhalerao:2015} & --- \\
SN~1987A & LMC & 35 & CC  & --- & \cite{Michael:2002,Zhekov:2005,Zhekov:2006,Dewey:2008,Zhekov:2009,Dewey:2012,Miceli:2019,Bray:2020,Ravi:2021,Sun:2021,Alp:2021} & \cite{Haberl:2006,Sturm:2010,Greco:2022} & --- \\
SNR~0509-67.5 & LMC & 310$^{+40}_{-30}$ & Ia  & --- & --- & \cite{Kosenko:2008,Williams:2011,Kosenko:2015} & --- \\
SNR~0519-69.0 & LMC & 600$\pm$200 & Ia  & --- & --- & \cite{Kosenko:2010,Williams:2011,Kosenko:2015} & --- \\
N103B & LMC & $\sim$800 & Ia  & --- & --- & \cite{vanderHeyden:2002,Yamaguchi:2021} & --- \\
SNR~0540-69.3 & LMC & $\sim$1200  & CC  &--- & --- & \cite{vanderHeyden:2001} & --- \\
N132D & LMC & $\sim$2500 & CC & \cite{Hwang:1993} & \cite{Canizares:2001} & \cite{Behar:2001,Suzuki:2020} & \cite{Hitomi:2017} \\
SNR~0506-68.0 & LMC & $\sim$4000  & CC  & --- & --- & \cite{Broersen:2011} & --- \\
DEM~L71 & LMC & $\sim$4400  & Ia  & --- & --- & \cite{vanderHeyden:2003} & --- \\
N49 & LMC & $\sim$4800 & CC & --- & --- & \cite{Amano:2020} & --- \\
SNR~0454-6713 & LMC & $\sim$8000 & Ia  & --- & --- & \cite{Seward:2021} & --- \\
SNR~0453.6-6829 & LMC & $\sim$13000  & CC & --- & --- & \cite{Haberl:2012,Koshiba:2022} & --- \\
SNR~0453-6655 & LMC & $\sim$60000  & CC & --- & --- & \cite{Seward:2018} & --- \\
1E~0102.2-7219 & SMC & $\sim$2000 & CC & --- & \cite{Canizares:2001,vanderHeyden:2004,Flanagan:2004} & \cite{Rasmussen:2001,Plucinskly:2017} & --- \\
SNR~0103-72.6 & SMC & $\sim$18000 & CC & --- & --- & \cite{vanderHeyden:2004} & --- \\
SN~1996cr & Circinus & 26 & CC & --- & \cite{Dwarkadas:2010,Dewey:2011,Quirola-Vasquez:2019} & --- & --- \\
SN~1978K & NGC~1313 & 44 & CC  & --- & --- & \cite{Chiba:2020} & --- \\
\noalign{\smallskip}\hline\noalign{\smallskip}
\end{tabular}
$^a$Distances to LMC, SMC, Circinus, and NGC~1313 are 50\,kpc \cite{Pietrzynski:2019}, 62\,kpc \cite{Graczyk:2020}, 3.8\,Mpc \cite{Koribalski:2004}, and 4.61\,Mpc \cite{Qing:2015}, respectively.\\
$^b$CC and Ia stand for core-collapse SN and Type Ia SN, respectively.\\
References for basic information about each SNR: Cygnus Loop \cite{Fesen:2021}, RX~J1713.7-3946 \cite{Ranasinghe:2022,Wang:1997}, Puppis~A \cite{Reynoso:2017,Ranasinghe:2022,Becker:2012}, SN~1006 \cite{Katsuda:2017}, RCW86 \cite{Ranasinghe:2022,Vink:2006}, Tycho's SNR \cite{Decourchelle:2017,Ruiz-Lapuente:2019}, Crab Nebula \cite{Blandford:2017,Fraser:2019}, Cas~A \cite{Koo:2017}, G296.1$-$0.5 \cite{Shan:2019,Gok:2012}, G21.5$-$0.9 \cite{Ranasinghe:2018,Bietenholz:2008}, Kepler's SNR \cite{Vink:2017}, G292.0$+$1.8 \cite{Ranasinghe:2022,Winkler:2009}, SN~1987A \cite{McCray:2016}, SNR~0509-67.5 \cite{Rest:2005,Hovey:2015}, SNR~0519$-$69.0 \cite{Rest:2005,Williams:2022}, N103B \cite{Rest:2005,Williams:2018}, SNR~0540-69.3 \cite{Lundqvist:2022}, N132D \cite{Vogt:2011}, SNR~0506-68.0 \cite{Hughes:2006,Broersen:2011}, DEM~L71 \cite{Ghavamian:2003}, N49 \cite{Park:2012}, SNR~0454-6713 \cite{Seward:2021}, SNR~0453.6-6829 \cite{Gaensler:2003a}, SNR~0453-6655 \cite{Seward:2018}, 1E~0102.2-7219 \cite{Finkelstein:2006,Xi:2019}, SNR~0103-72.6 \cite{Park:2003}, SN~1996cr \cite{Bauer:2008}, SN~1978K \cite{Ryder:1993}
\end{table}

\begin{figure}[htb]
\includegraphics[scale=.45]{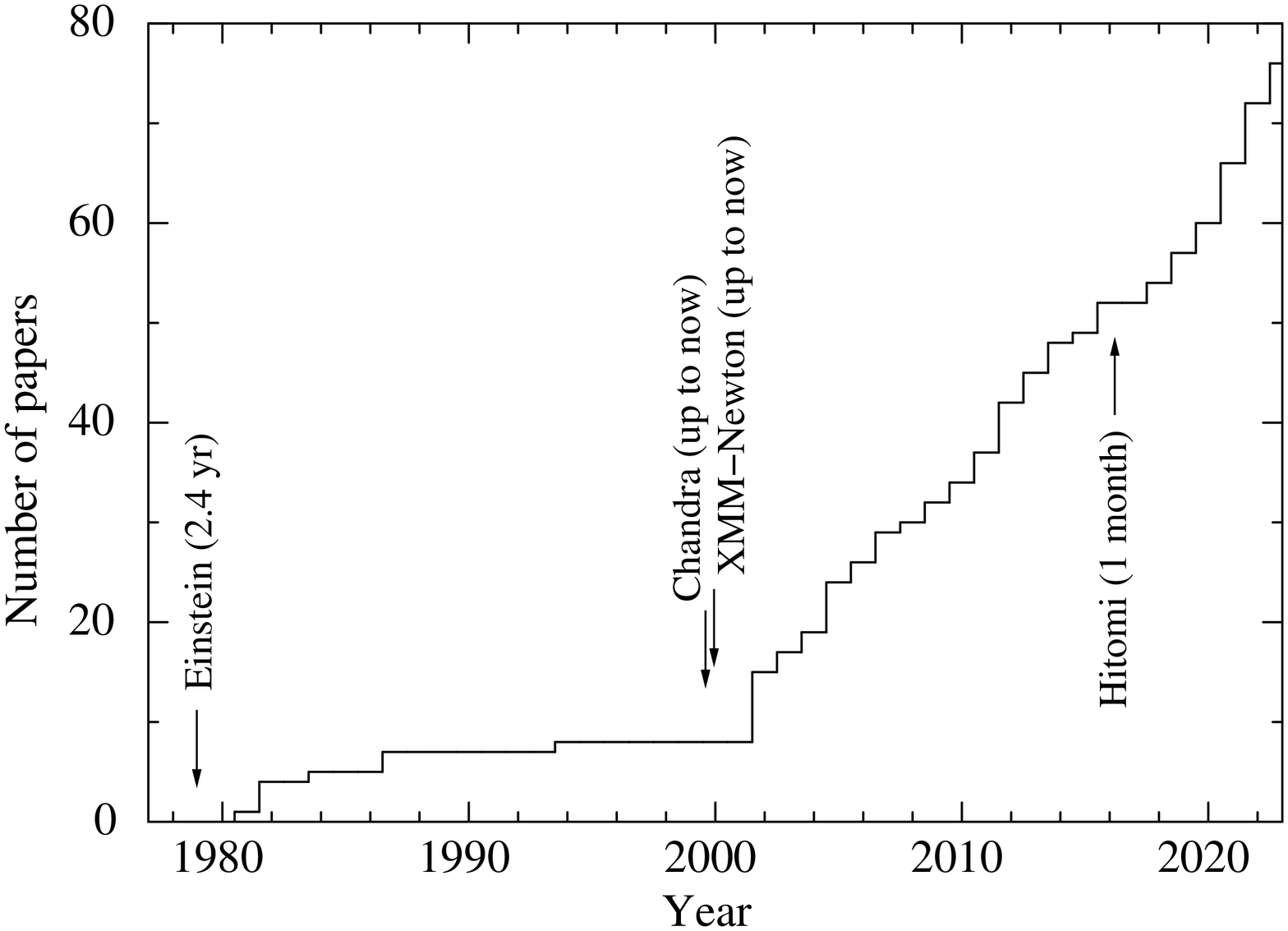}
%
%
\caption{Cumulative number of papers related to high-resolution X-ray spectroscopy of SNRs as a function of year.  The launches of the X-ray astronomy satellites, Einstein, Chandra, XMM-Newton, and Hitomi are indicated as arrows.}
\label{fig:papers_cum}       
\end{figure}

Arrays of low-temperature microcalorimeters are a relatively new technology for high-resolution X-ray spectroscopy, developed over recent few decades \cite{Akamatsu:2022,Sato:2023}.  It is a non-dispersive imaging spectrometer that measures the magnitude of the temperature pulse due to X-ray absorption to determine the X-ray energy.  The spectral resolution ($E/\Delta E = 100-1000$ depending roughly linearly on energy) is comparable with gratings.  A great advantage with the microcalorimeter array is that it works for diffuse sources as well as point-like sources.  The microcalorimeter (Soft X-ray Spectrometer, SXS: \cite{Kelley:2016}) aboard the Hitomi satellite \cite{Takahashi:2018} opened a new window to high-resolution X-ray spectroscopy for diffuse sources like SNRs.  Observations were successfully performed during its commissioning phase until the satellite was prematurely terminated by a series of abnormal events and mishaps triggered by the attitude control system.  Despite its short lifetime, Hitomi delivered a number of important results for a variety of astrophysical objects including SNRs.  

Table~\ref{tab:1} summarises SNRs in our Galaxy and Large and Small Magellanic Clouds (LMC and SMC) and two SNe in nearby galaxies, for which high-resolution X-ray spectroscopy has been successfully performed with publications.  Note that Einstein FPCS spectra for all observed sources including 11 SNRs are summarised in \cite{Lum:1992} --- a concise record of the FPCS observations.

Figure~\ref{fig:papers_cum} shows a cumulative number of papers as a function of year, from which we can see that the Einstein satellite initiated high-resolution X-ray spectroscopy of SNRs, and then this field has been vigorously developed with the advent of the Chandra and XMM-Newton satellites.  The soon-to-come X-ray astronomy satellite, X-Ray Imaging and Spectroscopy Mission (XRISM: \cite{Tashiro:2020}), will carry the X-ray microcalorimeter array (Resolve: \cite{Ishisaki:2022}) that is identical to that of Hitomi, and thus is anticipated to dramatically improve high-resolution X-ray spectroscopy of diffuse sources.  In the following subsections, we will present some key results in several scientific topics including plasma diagnostics (section~\ref{sec:plasma_diag}), new spectral features (section~\ref{sec:new_spec}), kinematics (section~\ref{sec:kinematics}), elemental abundances of the circumstellar medium, CSM (section~\ref{sec:csm}), and extinction features in X-ray spectra caused by either the intervening interstellar medium (ISM) or solar planets' atmospheres (section~\ref{sec:extinction}).

\section{Plasma Diagnostics}
\label{sec:plasma_diag}

SNR plasmas are usually in non-equilibrium (or transient) conditions.  The reason for this is twofold: low density and collisionless shock.  The density of the ISM is so low (typically 0.3\,cm$^{-3}$: \cite{Berkhuijsen:2008}) that shocks in the ISM are the so-called ``collisionless shocks", at which the ordered ion kinetic energy is dissipated into random thermal motions not by Coulomb collisions but through collisionless interactions involving magnetic fields.  Unlike the shocks formed in the Earth's atmosphere, which are mediated by molecular viscosity, the collisionless shocks in space lead to temperature non-equilibrium just behind the shock front.  This is naturally expected, because in the limit of true collisionless plasma, each particle is dissipated at the shock so that they will have temperatures proportional to their masses.  The post-shock temperature $T_{\rm a}$ for particle species $a$ with mass $m_{\rm a}$ is given by conservation of mass, momentum, and energy across the shock discontinuity as
\begin{eqnarray}
kT_{\rm a} = 3/16 m_{\rm a} v_{\rm s}^2
\label{eq:kT}
\end{eqnarray}
for shock velocity $v_{\rm s}$ \cite{Vink:2012}, assuming that the gas pressure is much greater than the magnetic pressure and the cosmic-ray particles' pressure.  Then, these different temperatures slowly approach equilibration via Coulomb collisions.  The timescales for electron-proton and electron-electron equilibrations are given by $t_{\rm ep} \sim$ 9800\,yr~$(n_{\rm p}/1\,{\rm cm^{-3}})^{-1}(kT/1\,{\rm keV})^{1.5}$ and  $t_{\rm ee} \sim$ 16\,yr~$(n_{\rm e}/1\,{\rm cm^{-3}})^{-1}(kT/1\,{\rm keV})^{1.5}$, respectively.

As a natural consequence of the temperature non-equilibration and evolution toward equilibration, SNR plasmas are expected to be in non-equilibrium ionisation (NEI) conditions.  Just after shock passages, post-shock plasmas are ionised only weakly, and then they are ionised slowly via Coulomb collisions mainly by free electrons  (protons contribute much less, because they are 1836 times more massive than electrons and they are 43 times slower than electrons).  The density is so low that it takes a long time for the plasma to reach ionisation equilibrium via Coulomb collisions.  As the temperature equilibration proceeds, the electrons are gradually heated.  Because the hotter electrons can ionise heavy elements to higher levels, the timescale for collisional ionisation equilibration should be longer than that of the temperature equilibration.  In this way, SNR plasmas are expected to be ionising during most of the SNR lifetime.  This kind of plasmas is known as the ``under-ionised plasma", and has been found in many SNRs.  Interestingly, the opposite NEI condition, in which the recombination rate is higher than the ionisation rate (``over-ionised plasma" or ``recombining plasma"), has been also found in some SNRs.  High-resolution X-ray spectroscopy offers excellent opportunities to study all kinds of the non-equilibrations.

\subsection{Kinetic Temperatures}
\label{sec:kinematic_T}

The timescale of the temperature equilibration is characterised by a product of the electron density multiplied by the time after the shock heating.  This parameter, $n_{\rm e}t$, is also known as the ionisation timescale, as it is sensitive to ionisation states.  A typical timescale for the plasma to be equilibrated is $n_{\rm e} t \sim$ 10$^{12}$\,cm$^{-3}$\,s.  Combined with the low density, it takes about 10$^{4}$\,yr for the SNR plasma to be relaxed.  This is longer than most of X-ray emitting SNRs.  Therefore, the temperature non-equilibration is expected in most of SNRs.  

\begin{figure}[htb]
\includegraphics[scale=.45]{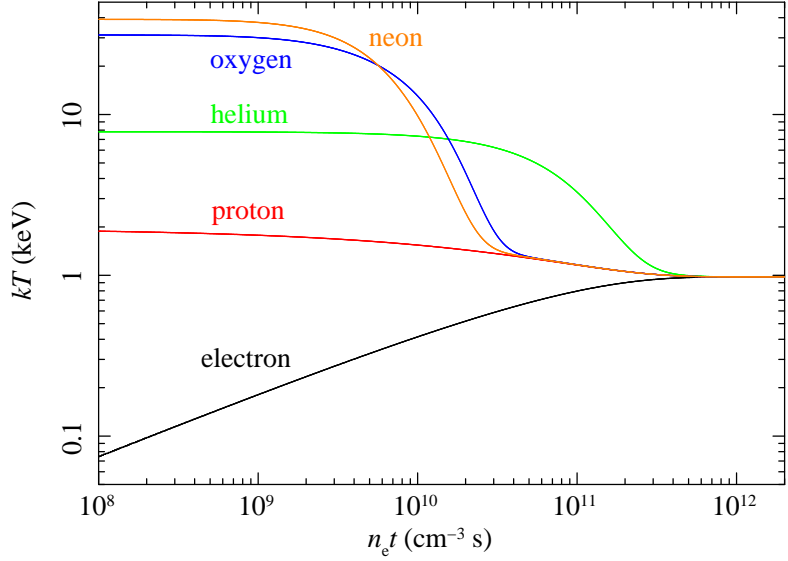}
%
%
\caption{Temperature evolution of shock-heated plasma as a function of ionisation timescale.  The shock speed is assumed to be 1000\,km\,s$^{-1}$.  The equilibration proceeds with Coulomb collisions.}
\label{fig:kT_evolution}       
\end{figure}

Figure~\ref{fig:kT_evolution} illustrates the effects of non-equilibration of temperatures at the shock front.  The equilibration time depends not only on the density, but also on the mass ratio between elements as well as the square of the charge number of the particles (see \cite{Vink:2012} for more details).  Therefore, the ion-ion equilibration proceeds faster than electron-ion equilibration.  Also, equilibration between heavy elements and protons proceeds faster than He-proton equilibration, due to the charge dependence.  

\begin{figure}[htb]
\includegraphics[scale=.45]{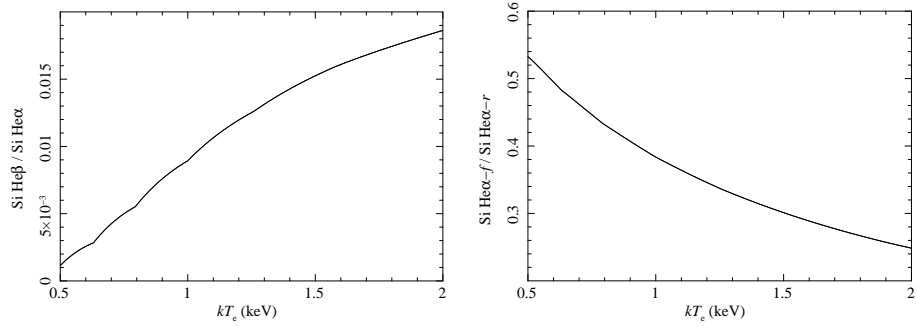}
%
%
\caption{Line intensity ratios of Si ions in an ionising plasma.  Left: Si He$\beta$ / Si He$\alpha$ as a function of electron temperature at a fixed ionisation timescale of 10$^{11}$\,cm$^{-3}$\,s.  Right: Same as left but for Si He$\alpha$-$f$ / Si He$\alpha$-$r$. }
\label{fig:line_diag1}       
\end{figure}

Electron temperatures can be estimated mainly by intensity ratios of lines from the same ion, such as He-like K$\alpha$ / He-like K$\beta$.  Practically, the flux ratio observed at the Earth is the emissivity ratio modified by a factor describing the relative interstellar absorption at the two lines with energies of $E_1$ and $E_2$, and is given by:
\begin{eqnarray}
\frac{F_{\rm 1}}{F_{\rm 2}} = \frac{\Omega_{1}}{\Omega_{2}} {\rm exp}[(E_{\rm 2} - E_{\rm 1})/kT_{\rm e}] {\rm exp}[(\sigma_{\rm E_2} - \sigma_{\rm E_1}) N_{\rm H} ],
\label{eq:Te}
\end{eqnarray}
where $\Omega$ is the effective collision strength or oscillator strength, $\sigma_{\rm E}$ gives the photo-electric absorption cross section at energy $E$, and $N_{\rm H}$ is the hydrogen column density for the intervening material.  Because oscillator strengths, line energies, and the cross sections of the intervening material are given in the literature, we can estimate the electron temperature by measuring the line flux ratio and $N_{\rm H}$.  Fig.~\ref{fig:line_diag1} left clearly demonstrates that the Si He$\beta$ / Si He$\alpha$ ratio strongly depends on the electron temperature.

Another useful clue to inferring the electron temperature is the forbidden-to-resonance ($f/r$) ratio in He-like ions.  The G($\equiv (f+i)/r$) ratio has been often used for this temperature diagnostics, but it can be usually approximated to the $f/r$ ratio, because the $i$ lines are usually much weaker than $f$ and $r$ lines.  The collisional excitation rates have different temperature dependence between $f$ and $r$ lines; the intensity of the $r$ line increases more rapidly with the temperature than the $f$ line \cite{Porquet:2010}.  Therefore, the $f/r$ ratio decreases with increasing temperature, as can be seen in Fig.~\ref{fig:line_diag1} right.  These line ratios ($\alpha$/$\beta$, $\beta$/$\gamma$, and $f$/$r$) have been measured for many SNRs, providing us with their electron temperatures.  

In reality, it is not rare that the temperatures from $f/r$ ratios are inconsistent with those from $\beta$/$\alpha$ ratios or global fittings \cite{Rutherford:2013,Suzuki:2020}.  This could be explained by contaminations of additional emission processes that will be described in Section~\ref{sec:new_spec}.  Therefore, we should keep in mind that electron temperatures estimated from $f/r$ ratios (or G ratios) are subject to some systematic uncertainties for the moment.

Ion temperatures can be estimated from line broadening which is of the order of 1\,eV and thus can be accessible only with high-resolution X-ray spectroscopy ($E/\Delta E \gtrsim 100$).  The velocity distribution of thermal particles (ions) follows the Maxwellian distribution:
\begin{eqnarray}
f(v) = \sqrt{m/2\pi kT}{\rm exp}(-mv^2/2kT).
\end{eqnarray}
The lines emitted from ions are Doppler shifted following this velocity distribution, so that the line profile can be represented by a Gaussian distribution with
\begin{eqnarray}
\sigma = E_0/c \sqrt{kT/m},
\end{eqnarray}
where $E_0$ is the line energy and $c$ is the speed of light.  Combining this equation with the relation between shock speed and post-shock temperature (Eq.~\ref{eq:Te}), we can estimate an expected line width to be $\sigma = 1.4\,{\rm eV} (E_0/1\,{\rm keV}) (v_{\rm s}/1000\,{\rm km\,s}^{-1})$.  Because this width is much smaller than the spectral resolution of most of past instruments aboard X-ray astronomy satellites, there are only a few successful measurements of ion temperatures.  X-ray detections of significant line broadenings were limited only for SN~1006's northwestern knot \cite{Vink:2003,Broersen:2013} and SN~1987A \cite{Miceli:2019}.  In both cases, ion temperatures were found to be much higher than electron temperatures, evidencing the non-equilibration of temperatures in collisionless shocks.  Moreover, \cite{Miceli:2019} reported that ion temperatures increase with the ion mass from proton to Fe as $T_{\rm i}/T_{\rm p} = (0.90\pm0.12) \times m_{\rm a}$.  Therefore, the ion post-shock temperature is consistent with being mass-proportional, which is expected for a truly collisionless shock.

Even upper limits on the line widths can give us useful information.  The RGS spectrum taken from fast-moving ejecta knots in Puppis~A revealed line broadening at O Ly$\alpha$ to be $\sigma \lesssim 0.9$\,eV, indicating an O temperature of $\lesssim 30$\,keV \cite{Katsuda:2013}.  This O temperature, combined with the electron temperature and ionisation timescale, led to a conclusion that the ejecta knots were heated by a (reverse) shock with a velocity of 600--1200\,km\,s$^{-1}$.

\subsection{Ionisation States}

\begin{figure}[htb]
\includegraphics[scale=.45]{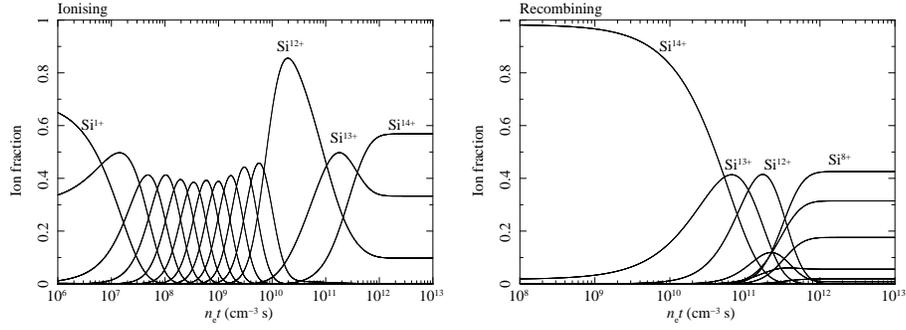}
%
%
\caption{Left: Ion population of Si in an ionising plasma as a function of ionisation timescale.  The electron temperature is assumed to be constant at 1.5\,keV.  This figure is created with SPEX.  Right: Same as left, but for a recombining plasma.  The initial and final temperatures are assumed to be 10\,keV and 0.1\,keV, respectively.}
\label{fig:ion_frac}       
\end{figure}

The effect of the NEI condition is illustrated in Fig.~\ref{fig:ion_frac}, showing how the ionisation state (of Si ions as a representative element) in the NEI plasma proceeds with the ionisation timescale.   The left and right panels show under-ionised and over-ionised plasmas, respectively.  For each case, ionisation or recombination proceeds with time.  Both plots are generated using the SPEX code \cite{Kaastra:1996}.  We assume a constant temperature of 1.5\,keV for the under-ionised case, whereas we assume the initial temperature of 10\,keV and the final equilibrated temperature of 0.1\,keV for the over-ionised case.  In both cases, the collisional equilibration is reached when $n_{\rm e} t \sim$ 10$^{12}$\,cm$^{-3}$\,s.  We also note that He-like ions (Si$^{12+}$) are abundant for a long time due to their closed shell ground state.  This is why we often see the most prominent X-ray spectral lines of He-like ions among other ions in hot plasmas from a variety of astrophysical sources.

\begin{figure}[htb]
\includegraphics[scale=.45]{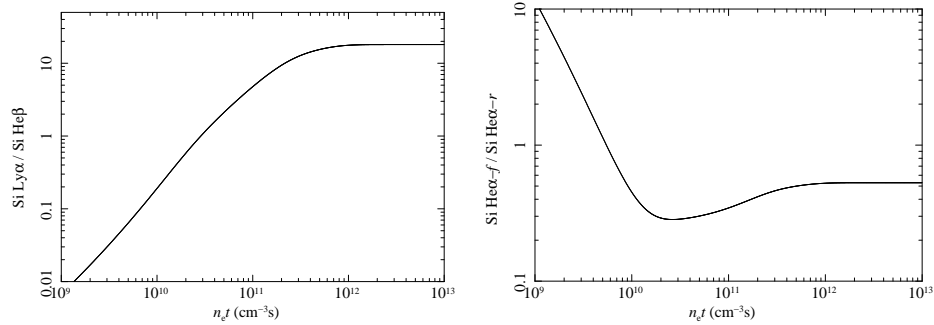}
%
%
\caption{Line intensity ratios of Si ions in an ionising plasma.  Left: Si Ly$\alpha$ / Si He$\beta$ as a function of ionisation timescale at a fixed electron temperature of 1.5\,keV.  This figure is created with SPEX.  Right: Same as left but for Si He$\alpha$-$f$ / Si He$\alpha$-$r$. }
\label{fig:line_diag2}       
\end{figure}

High-resolution X-ray spectroscopy allows one to measure the degree of the NEI condition or the ionisation timescale of a plasma, based on a combination of the ion fraction and the electron temperature.  Historically, the NEI conditions in SNR plasmas were first found with the Einstein FPCS.  The ion fraction can be estimated by line intensity ratios of different transitions involving different ionisation states of the same element.  Similar to Eq.~\ref{eq:Te}, we can derive  the ion fraction from the equation:
\begin{eqnarray}
\frac{F_{\rm 1}}{F_{\rm 2}} = \frac{n_{\rm Zz}}{n_{\rm Zx}} \frac{\Omega_{1}}{\Omega_{2}} {\rm exp}[(E_{\rm 2} - E_{\rm 1})/kT_{\rm e}] {\rm exp}[(\sigma_{\rm E_2} - \sigma_{\rm E_1}) N_{\rm H} ],
\label{eq:ion_frac}
\end{eqnarray}
where $n_{\rm Zz}$ and $n_{\rm Zx}$ are number densities of ions Z$^{\rm Z+}$ and Z$^{\rm X+}$.  The cleanest flux ratio is Ly$\alpha$ ($1s^2-1s2p$) / He$\beta$ ($1s^2-1s3p$), because these two lines are located so closely that the exponential terms in Eq.~\ref{eq:ion_frac} are of order 1.  Therefore, it is straightforward to retrieve the ion fraction between He-like and H-like ions.  Figure~\ref{fig:line_diag2} left displays a flux ratio between Si Ly$\alpha$ and Si He$\beta$ as a function of ionisation timescale, where the flux ratio is calculated at an assumed constant temperature of 1.5\,keV with the SPEX code.  The Ly$\alpha$/He$\beta$ ratio increases monotonically with increasing ionisation timescale, as expected.

\begin{figure}[htb]
\includegraphics[scale=.5]{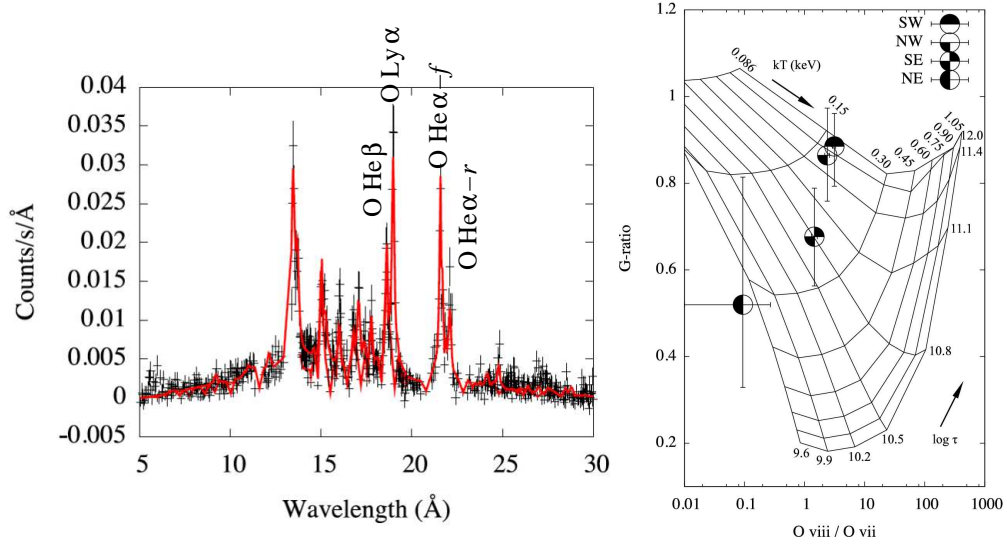}
%
%
\caption{Left: RGS spectrum of the northwestern rim of RCW~86.  The solid red line represents the best-fit model consisting of an absorbed, two NEI plus one power-law components.  Right: Grid of temperature and ionisation timescale created with SPEX.  The line ratios observed at the different four regions of the SNR are plotted by partially-filled circles as indicated in the right-upper corner.  These figures are taken from \cite{Broersen:2014}.}
\label{fig:rcw86}       
\end{figure}

If we combine the ion fraction with an electron temperature, we can estimate the ionisation timescale of the plasma.  As was explained in Section~\ref{sec:kinematic_T}, the electron temperature can be estimated from either the line intensity ratio between different transitions in the same ion (e.g., He-like K$\beta$ / He-like K$\beta$), or the G-ratio of He-like K$\alpha$ lines.  Figure~\ref{fig:rcw86} demonstrates such an example, where the O Ly$\alpha$ to O He$\beta$ ratio and the G-ratio are used to estimate ion fractions and electron temperatures for four regions in a young Galactic SNR, RCW~86 \cite{Broersen:2014}.  The results show that two regions are far from ionisation equilibrium, and the other two regions are consistent with collisional equilibrium.  In the same manner, NEI conditions have been found in many SNRs including Puppis~A \cite{Winkler:1981a,Winkler:1981b}, Cygnus Loop \cite{Vedder:1986}, and 1E~0102.2-7219 \cite{Rasmussen:2001}, and so on.  We remark that moderate spectral resolution ($E/\Delta E \sim 20$) is sufficient to resolve K$\alpha$ and K$\beta$ lines from He-like and H-like ions.  Therefore, non-dispersive X-ray CCDs and even gas scintillation proportional counters successfully identified NEI conditions in many SNRs \cite{Tsunemi:1986,Miyata:1994,Yamaguchi:2008}.  

Alternatively, flux ratios of the K$\alpha$ lines in He-like ions can be used to estimate the NEI condition.  Detailed descriptions of He$\alpha$ diagnostics can be found in \cite{Porquet:2010}.  We here focus on one powerful diagnostic parameter, i.e., the $f$/$r$ intensity ratio as shown in Fig.~\ref{fig:line_diag2} right.  The resonance lines are emitted at the moment of collisional excitations mainly due to electrons.  On the other hand, collisional excitations from the ground level to the triplet level ($^3$P$_{0,1,2}$ and $^3$S$_1$; see e.g., Fig.~1 of \cite{Porquet:2010}) are much less efficient according to the selection rule.  Thus, the intercombination and forbidden lines are emitted after either the inner-shell ionisations of Li-like ions or recombinations of the H-like ions.  These processes favor the forbidden triplet levels because of its high statistical weight.  Therefore, at the very low-ionisation condition in which Li-like ions are more populated than He-like and H-like ions, we expect a high $f/r$ ratio.  Also, for (equilibrated) plasmas with full of H-like ions, the He$\alpha$ lines are dominantly produced by recombination processes.  Because recombination processes (both radiative and di-electronic recombinations) favor the triplet level due to the high statistical weight, we expect an intense He$\alpha$-$f$ line (for low density plasmas), leading to a high $f/r$ ratio.  In contrast, when the abundance of He-like ions becomes dominant, with little fractions of Li-like and H-like ions, we expect a low $f/r$ ratio.  Such a trend can be readily seen in Fig.~\ref{fig:line_diag2} right showing $f/r$ ratios in He-like Si ions as a function of ionisation timescale.  The low $f/r$ ratios observed in many SNRs are considered as such a mild NEI condition: Puppis~A \cite{Winkler:1981a,Winkler:1981b}, Cygnus Loop \cite{Vedder:1986}, and 1E~0102.2-7219 \cite{Rasmussen:2001}.

On the other hand, an enhanced O VII $f/r$ ratio was found in a middle-aged SNRs, DEM~L71 and 0506-68 in the LMC.  \cite{vanderHeyden:2003,Broersen:2011} argued that the high $f/r$ ratio is most naturally explained by a contamination of recombining plasmas, given that the X-ray spectrum can be best explained by a cooling (or recombining) plasma model.  However, radiative recombination edges, which are strong signatures of rapidly cooling plasmas found in mixed-morphology SNRs like IC 443 \cite{Yamaguchi:2009} and W49B \cite{Ozawa:2009}, were not observed in the two LMC SNRs.  Therefore, other possibilities like resonance line scattering are still left as a viable mechanism to create the high $f/r$ ratio.

\section{New Spectral Features}
\label{sec:new_spec}

Recent high-resolution X-ray spectroscopy of SNR plasmas has been revealing that line intensity ratios are often inconsistent with a conventional thermal emission models.  Two main possibilities to explain the anomalous line ratios are effects of charge exchange (CX) and resonance scattering (RS).  Both effects were predicted to be present in SNRs three decades ago \cite{Wise:1989,Lallement:2004,Kaastra:1995}, but are not yet established firmly.  In another chapter of this book, \cite{Gu:2023} gave a thorough review on the current status of CX X-rays on three levels: theoretical calculations of the cross sections, laboratory measurements of reaction rates and line strengths, and the observational constraints using celestial objects.  

The CX process simply transfers (an) electron(s) from one atom to another.  No photon is created via this electron transfer itself, but the electron maintains roughly the same binding energy in the process of moving, so the recipient ion is usually left in a highly-excited state (typically $n = 5$ to $n=10$) that then will be stabilised by radiative cascades (see \cite{Smith:2014} for detailed explanations on CX processes).  From an observational perspective, it is important to note that (1) CX into a He-like triplet state will lead to enhanced $f$ and $i$ lines (because almost any exchange into a triplet state results in such lines), and (2) CX into the He-like singlet state will lead to enhanced (relative to a purely collisional model in the ground state) high-$n$ resonance transitions.  These features are similarly expected in the recombining plasmas, because X-rays are emitted via radiative cascade in both processes.  One remarkable difference is that radiative recombination edges, which are created when free electrons move to ions, should be seen in recombining plasmas, but not in CX processes.  

CX emission in SNRs is thought to play an enhanced role where neutrals are mixed with shocked hot gas.  \cite{Wise:1989} first performed detailed calculations of CX-induced X-ray emission in a SNR.  They found that CX emission generally contributes only 10$^{-3}$ to 10$^{-5}$ of the collisionally excited lines in the entire SNR.  On the other hand, \cite{Lallement:2004} later examined projected emission profiles for both CX and thermal emission in SNRs, and noted that CX X-ray emission may be comparable with thermal emission in thin layers at the SNR edge.  It is also pointed out that the relative importance of CX X-ray emission to thermal emission is proportional to a quantity, $n_{\rm c} v_{\rm r} n_{\rm e}^{-2}$ , with $n_{\rm c}$ being the cloud density, $v_{\rm r}$ the relative velocity between neutrals and ions, and $n_{\rm e}$ the electron density of the hot plasma.  Thus, the higher the density contrast ($n_{\rm c}$/$n_{\rm e}$), the stronger the presence of CX X-ray emission would become.

\begin{figure}[htb]
\includegraphics[scale=0.3]{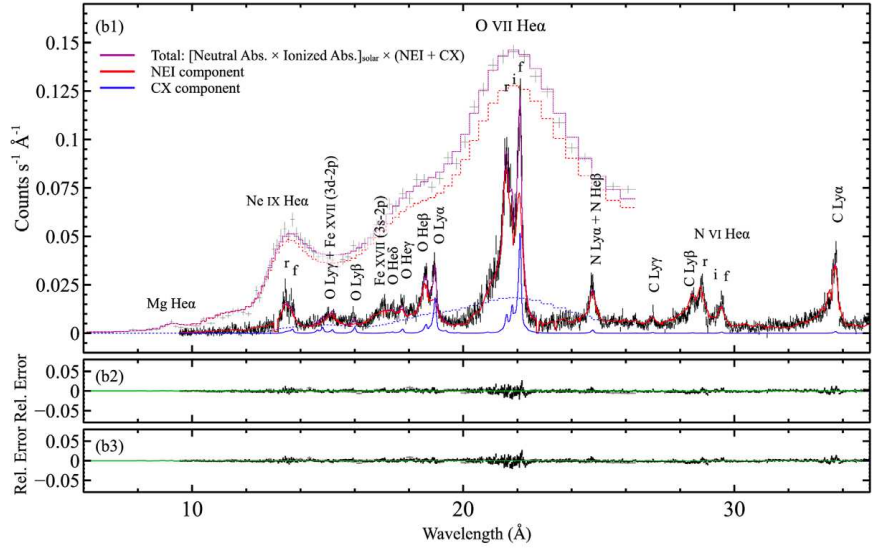}
%
%
\caption{RGS1$+$2 (black) and pn (grey) spectra taken from the southwestern knot in the Cygnus Loop.  The best-fit model represents an absorbed, NEI (red) $+$ CX (blue) components, where the absorption model consists of both neutral and ionised components.  The residuals shown in panels (b2) and (b3) are for different absorption models [neutral abs. $\times$ ionised abs.]$_{\rm solar~abundance}$ and [neutral~abs.]$_{\rm solar~abundance}$ $\times$ [ionised~abs.]$_{\rm cygnus~loop}$, respectively.  This figure is taken from \cite{Uchida:2019}.}
\label{fig:cygnus_rgs}       
\end{figure}

A strong emission feature at $\sim$0.7\,keV in parts of outermost rims of the Cygnus Loop was best interpreted as the O He$\gamma + \delta + \epsilon$ arising from CX between neutrals and H-like O ions \cite{Katsuda:2011}.  Later, \cite{Uchida:2019} discovered a high $f/r$ ratio of O He$\alpha$ at the southwestern knot of the Cygnus Loop, using the XMM-Newton RGS as shown in Fig.~\ref{fig:cygnus_rgs}.  This anomalous $f/r$ ratio is best interpreted as a result of a contribution from CX emission.  High $f/r$ ratios were also found in some other SNRs including Puppis~A \cite{Katsuda:2012}, G296.1$-$0.5 \cite{Tanaka:2022}, and SNR~0453.6$-$6829 \cite{Koshiba:2022}, suggesting the presence of CX X-ray emission.  However, other possibilities such as resonance scattering or inner-shell ionization are not ruled out.  

The importance of RS in SNR plasmas was first pointed out by \cite{Kaastra:1995}, who argued that the optical depth for resonance lines cannot be assumed to be negligibly small for bright SNRs.  If there is a sufficient ion column density along a particular line of sight, resonance line photons can be scattered out of the line of sight to appear at another location.  The SNR rims are promising sites where RS effects can be significant, because lines of sight are usually long enough there and thus large optical depths are expected.  Observationally, lines with large oscillator strengths (e.g., He$\alpha$-$r$ and Ly$\alpha$) will be suppressed at SNR rims, whereas those with small oscillator strengths (e.g., He$\alpha$-$f$ and He$\beta$) will not be affected.

Anomalously high $f/r$ ratios of O He$\alpha$ observed in some SNRs like DEM~L71 \cite{vanderHeyden:2003} and N49 \cite{Amano:2020} in the LMC were best interpreted as the suppression of $r$ lines due to the effect of RS.  Using the RGS data, \cite{vanderHeyden:2003} revealed spatial variations in the $f/r$ ratio of O He$\alpha$, and raised two possibilities for the high $f/r$ ratio in DEM~L71: (1) dramatically cooling and recombining plasmas in some regions and (2) RS effects.  Strong support for the RS effect is provided by the O line profiles that the distribution of O Ly$\alpha$ and O He$\alpha$-$f$ are similar with each other, while O He$\alpha$-$r$ is quite distinct, suggesting that the high $f/r$ ratio is brought about by a reduction in the O He$\alpha$-$r$ rather than an enhancement of the O He$\alpha$-$f$.  \cite{Amano:2020} revealed in N49 that O VIII Ly$\beta$/$\alpha$ and Fe XVII ($3s-2p$)/($3d-2p$) ratios can be represented by thermal NEI emission model modified by the RS effect.  Also, in an effort to interpret the high $f/r$ ratio in the southwestern knot in the Cygnus Loop obtained with the XMM-Newton RGS, \cite{Uchida:2019} pointed out a possibility that the global high-resolution X-ray spectrum can be significantly better fitted by introducing ionised absorber.  The origin of the ionised absorber remains unclear, but the most plausible explanation seems a local self-absorption, i.e., the effect of RS (cf.\ Fig.~\ref{fig:cygnus_rgs} (b3)). 

Yet another possibility to explain the anomalously high $f/r$ ratio is the inner-shell ionisation of Li-like ions \cite{Smith:2014}.  This process creates the excited $1s2s$ state, and hence can significantly contribute to the He$\alpha$-$f$ line, but not to the He$\alpha$-$i$ line.  This is in stark contrast to the cascading processes such as CX and recombination in which both $f$ and $i$ lines are enhanced.  Therefore, the relative intensities of He$\alpha$-$f$, $i$, and $r$ may be a useful diagnostic tool to fully understand the X-ray emission processes in SNR plasmas.

\section{Kinematics of SN Ejecta and Circumstellar Medium}
\label{sec:kinematics}

Core-collapse SNe that occurred in nearby galaxies often emit intense X-rays within the first 1000 days, due to the interaction between SN ejecta and the CSM blown by the progenitor star.  Among many extragalactic SNe detected in X-rays \cite{Ross:2017,Chandra:2018}, SN~1987A and SN~1996cr are outstanding objects as they show increasing X-ray fluxes with time.  In addition, they are the only two sources for which high-resolution X-ray spectroscopy was successfully performed with Chandra and XMM-Newton.

A large number of SNRs with ages $\gtrsim$ 100\,yr are visible in X-rays (regardless of the SN type) in our Galaxy, the LMC, and the SMC.  High-resolution X-ray spectroscopy has been performed either for the entire regions of bright and small SNRs in the L/SMCs, or for small knots and filaments embedded in largely-extended Galactic SNRs.  Below, we will concentrate on kinematics of SNe and SNRs derived from high-resolution X-ray spectroscopy.  This section is a concise update of a previous review of X-ray measured kinematics of SNRs by \cite{Dewey:2010}.

\subsection{Global Ejecta Structures in Extragalactic SNe and SNRs}
\label{sec:global_ejecta}

\begin{figure}[htb]
\includegraphics[scale=0.5]{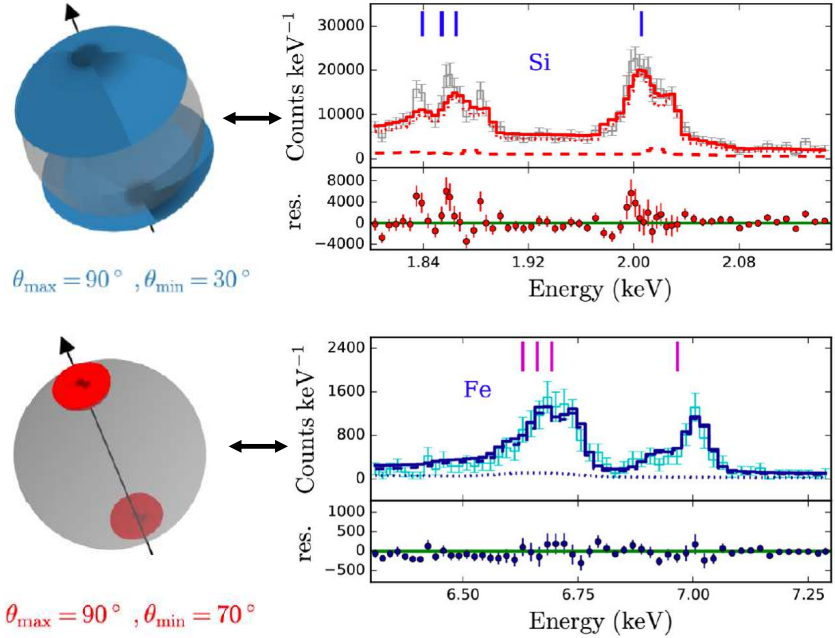}
\caption{Left-upper and left-lower panels show expanding shock structure geometries for a 60$^\circ$ wide polar cap and a 20$^\circ$ wide polar cap, respectively.  Right-upper and right-lower panels show close-up spectra for Si K lines and Fe K lines, respectively.  Both of them are taken from the Chandra HETG observation in 2009.  The best-fit models assume polar cap geometries with opening angles of 30$^\circ$ and 60$^\circ$ for Si and Fe, respectively.  This figure is taken from \cite{Quirola-Vasquez:2019} with authors' permission.}
\label{fig:96cr}       
\end{figure}

The Chandra HETG presented spectacular X-ray spectra from SN~1996cr in the Circinus Galaxy, revealing velocity distributions of Ne, Mg, Si, S, and Fe from line emission profiles \cite{Quirola-Vasquez:2019}.  The X-ray spectra are well represented by thermal emission arising from CSM-ejecta interactions undergoing an obscured, shell-like expansion.  The line profiles require that shocked regions are distinctly not spherically symmetric, suggesting a polar geometry with two distinct opening angle configurations and internal obscuration.  The lines from Si and other elements except for Fe can be best represented by a mildly absorbed ($N_{\rm H} \sim 2\times$10$^{21}$\,cm$^{-2}$), cooler ($kT_{\rm e}$$\sim$2\,keV) plasma with high Ne, Mg, Si, and S abundances associated with a wide polar interaction region (half-opening angle of 58$^\circ\pm4^\circ$) as illustrated in upper panels of Fig.~\ref{fig:96cr}.  On the other hand, lines from Fe can be explained by a moderately absorbed ($N_{\rm H} \sim 2\times$10$^{22}$\,cm$^{-2}$), hotter ($kT_{\rm e}$$>$20\,keV) plasma with high Fe abundances and strong internal obscuration associated with a narrow polar interaction region (half-opening angle of 20$^\circ\pm5^\circ$) as in lower panels of Fig.~\ref{fig:96cr}.  \cite{Quirola-Vasquez:2019} argued that the cooler (Si) and hotter (Fe) components are associated with forward and reflected shocks, and thus originate from the dense CSM and ejecta, respectively.  Interestingly, the X-ray spectrum shows temporal enhancements of metal abundances between 2000 and 2018.  This implies metal enhancements from shocked ejecta material possibly due to the fingers of Rayleigh-Taylor instabilities for the forward shock component or just that more Fe-rich ejecta are being shocked as the shock moves deeper inwards for the reflected shock component.  Unfortunately, the XMM-Newton RGS data are not useful for this source because of the difficultly in separating SN~1996cr's emission from the bright extended emission associated with the AGN and circumnuclear star formation.

X-ray emission from SN~1987A has been dominated by shock interactions with the dense inner ring, also known as the equatorial ring (ER).  The line profiles obtained with the Chandra LETG enabled us to reveal the kinematics of the ER.  The optical inner ring has an inclination angle $i = 44^\circ-45^\circ$ with the near side to the north, so that the north side of the ring is blueshifted, whereas the south side is redshifted.   Then, the dispersion axis of the LETG was set along the north-south direction, with the negative ($m = -1$) arm pointing to the north.  In this case, the northern (southern) side of the ring will be displaced to the shorter (longer) wavelength along the negative arm, hence both sides approach with each other.  On the other hand, both sides will be stretched in the positive arm.  Given that the northern and southern sides cannot be resolved even with Chandra's superior spatial resolution, the line profiles include both sides, yielding apparently single lines.  Therefore, the lines are expected to be narrower in the negative arm than in the positive arm.  This effect was indeed observed in the LETG spectrum (Fig.~1 right in \cite{Zhekov:2005}).

The same technique was applied to infer the three-dimensional geometry of the SNR 1E~0102.2-7219 \cite{Flanagan:2004}.  The dispersed images of Ne Ly$\alpha$ show clear distortions relative to the zeroth order image.  The $-1$ and $+1$ order Ne Ly$\alpha$ images are different from each other, with a sharp $-1$ order and a broad $+1$ order.  This can be explained by a cylindrical or thick-ring distribution of the ejecta, with velocities of order 1000\,km\,s$^{-1}$.  

Doppler effects of expanding ejecta were measured for some other SNRs: N132D, SNR~0509-67.5, SNR~0519-69.0, and N103B in the LMC.  N132D is a relatively young (age $\sim$2500\,yr: \cite{Vogt:2011}), core-collapse SNR.  The Einstein FPCS measured the width of O Ly$\alpha$ in excess of the expected width from the spectral and spatial resolution of the detector to be 1000--2000\,km\,s$^{-1}$ \cite{Hwang:1993}.  Using a short 3.7\,ks observation with the Hitomi SXS (note that the exposure of 3.7\,ks includes times when the remnant was partially outside the field of view of the SXS), \cite{Hitomi:2017} successfully detected K-shell lines from highly-ionised S and Fe ions.  The widths were measured to be $\sigma \sim$500\,km\,s$^{-1}$ for both S K and Fe K lines.  This is consistent with the result from the Einstein satellite (if we assume that the width reported in \cite{Hwang:1993} is FWHM).  More recently, \cite{Suzuki:2020} measured velocity dispersions averaged over the entire X-ray spectrum with the XMM-Newton RGS to be $\sigma \sim 500$\,km\,s$^{-1}$, which is consistent with past observations.

Much faster expansions were found for young Type Ia SNRs~0509-67.5 and 0519-69.0.  Using the O K and Fe L lines measured with the RGS, \cite{Kosenko:2008,Kosenko:2010} measured global velocity broadenings to be $\sigma \sim 4900$\,km\,s$^{-1}$ and $\sigma \sim 1900$\,km\,s$^{-1}$ for 0509-67.5 and 0519-69.0, respectively.  N103B is also a young Type Ia SNR, but the RGS spectrum does not require any additional line broadening beyond the spatial contribution, suggesting an upper limit of $\sigma < 350$\,km\,s$^{-1}$ \cite{vanderHeyden:2002}.  The diversity of the line broadenings found in these three Type Ia SNRs can be interpreted as different amounts of the ambient medium; the denser (more massive) the ambient medium, the stronger the deceleration of the expansion velocity.

\subsection{Kinematics of Galactic SNRs}

Galactic SNRs allow us to explore detailed three-dimensional structures inside the remnant, if they have bright and compact knots that are suitable for spectroscopy with slitless gratings aboard X-ray astronomy satellites.  So far, high-resolution X-ray spectra have been obtained for SN ejecta and the CSM in Cas~A, G292.0+1.8, Puppis~A, Kepler's SNR, and Tycho's SNR.  Below, we will briefly review individual cases.  

\begin{figure}[htb]
\includegraphics[scale=0.4]{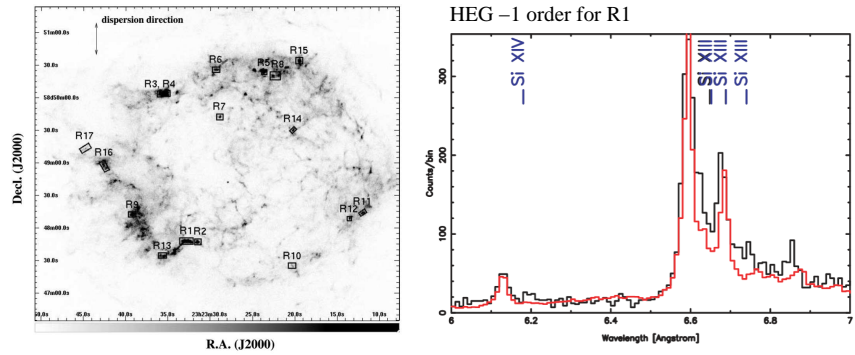}
\caption{Left: Chandra X-ray image of Cas~A with regions for the high-resolution spectroscopy with the Chandra HETG.  Right: Example Chandra HETG spectrum of the Si K band taken from the region R1 indicated in the left panel.  The best-fit model is plotted with a red line.  The nominal positions of the Si lines are marked.  These figures are taken from \cite{Lazendic:2006}.  \textcircled{C} AAS.  Reproduced with permission.}
\label{fig:CasA}       
\end{figure}

{\it Cas~A}: One of the best-studied, young core-collapse SNR.  He$\alpha$ and Ly$\alpha$ lines' profiles of Si and S were obtained for the whole SNR with the Einstein FPCS, with the aperture of 6$^\prime$.  The orientation of the spacecraft during the FPCS observation was such that the bright lobes in the northwest and southeast were spatially resolved from one another.  The X-ray spectrum of the northwestern portion was found to be redshifted with respect to that of the southeast, with a mean velocity difference between the two regions of 1820$\pm$290\,km\,s$^{-1}$.  The velocity asymmetry has been confirmed by following observations with non-dispersive CCDs aboard ASCA \cite{Holt:1994}, Chandra \cite{Hwang:2001}, and XMM-Newton \cite{Willingale:2002}.  Later, using the Chandra HETG, \cite{Lazendic:2006} analysed line emission dominated by Si and S ions for many bright, narrow regions of Cas~A to examine their kinematics as well as plasma non-equilibrations.  The selected regions are shown in Fig.~\ref{fig:CasA} left.  Unambiguous Doppler shifts for these selected regions were found.  Figure~\ref{fig:CasA} right shows an example HETG spectrum taken from the R1 knot indicated in Fig.~\ref{fig:CasA} left.  The lines are significantly blueshifted as shown in Fig.~\ref{fig:CasA} right, with the best-fit blueshift of 2600\,km\,s$^{-1}$.  Overall, the knots in the southeastern portion mostly show blueshifted velocities up to $-$2500\,km\,s$^{-1}$, whereas knots in the northwestern side mostly show redshifted velocities up to $+$4000\,km\,s$^{-1}$, which is in general agreement with past observations with lower spatial and spectral resolutions.  More recently, \cite{Rutherford:2013} investigated temporal evolution using Chandra HETG observations spanning a 10\,yr baseline.  Due to the low ionisation timescale and the high electron density derived from the X-ray analysis, the small features were expected to show significant amounts of plasma evolution during the 10\,yr baseline.  However, most of them showed insignificant time evolution, suggesting a much lower electron density for the Si-rich plasmas and thus a longer timescale for evolution of the plasma.

{\it G292.0+1.8}: Another relatively young core-collapse SNR.  The Chandra HETG revealed Doppler shifts in emission lines from metal-rich ejecta, providing three-dimensional structures of clumpy ejecta material in the remnant \cite{Bhalerao:2015}.  The distribution of ejecta knots in velocity versus projected-radius space suggests an expanding ejecta shell with a projected angular thickness of $\sim$90$^{\prime\prime}$ (corresponding to $\sim$3 pc at a distance of 6\,kpc).  Based on this geometrical distribution of the ejecta knots, the location of the reverse shock is estimated approximately at the distance of $\sim$4 pc from the center of the SNR, putting it in close proximity to the outer boundary of the radio pulsar wind nebula.  It should be noted that \cite{Vink:2004} presented high-resolution X-ray spectrum with the XMM-Newton RGS for the central belt-like feature across the remnant.  No significant line broadening is indicated by the O Ly$\alpha$ with $\sigma < 730$\,km\,s$^{-1}$, but significant line broadening seems to be present for the Ne Ly$\alpha$ with $\sigma \sim 1500$\,km\,s$^{-1}$.  Given that most of O and Ne lines are associated with the ISM/CSM and SN ejecta, respectively, this kinematic result suggests that the blastwave has decelerated considerably, whereas some of the ejecta are still moving with a high velocity.

{\it Puppis~A}: One of the three ``O-rich" SNRs in our Galaxy, with others being Cas~A and G292.0+1.8.  The X-ray emission is dominated by the swept-up CSM and/or ISM, whereas some ejecta features have been identified in X-rays \cite{Hwang:2008,Katsuda:2008,Katsuda:2010}.  \cite{Katsuda:2013} observed O-rich ejecta knots and filaments with the RGS, finding that the knots located near the center of the remnant have a Doppler velocity of $\sim$1500\,km\,s$^{-1}$ blueward and the filament located in the eastern rim has a Doppler velocity of $\sim$650\,km\,s$^{-1}$ redward.  Given that they are SN ejecta, it is reasonable to assume that they were heated by reverse shocks.  In this case, $T_{\rm O}$, $T_{\rm e}$, and the ionisation timescale, combined with Coulomb equilibration after the shock heating, suggested that the free expansion speeds of both of these ejecta features were  around 2500--3000\,km\,s$^{-1}$. \footnote{Note that equation (3) in \cite{Katsuda:2013} should be better replaced to $v_{\rm fs} = (1-\gamma)/2 \times v_{\rm sg} +(1+\gamma)/2 \times v_{\rm ej}$ to take account of the motion of the unshocked ejecta knot.  This change has no impact on the interpretation related to the reverse-shock scenario.  I thank Pat Slane for pointing it out.}  Such a high speed is consistent with typical ejecta velocities found in the O-rich layer for Type IIb SN, which is a possible subtype of the SN explosion that produced Puppis~A \cite{Chevalier:2005}.

{\it Kepler's SNR}: The remnant of SN~1604 --- the most recent Galactic historical SN.  This SNR is a rare class of Type Ia SNRs that show interactions with the CSM.  The X-ray emission comes from both the CSM and SN ejecta that are associated with the forward and reverse shocks, respectively.  \cite{Sato:2017b} measured proper motions and Doppler velocities for 14 compact X-ray knots, using data from non-dispersive CCDs aboard Chandra.  They found high Doppler velocities of up to $\sim$10$^4$\,km\,s$^{-1}$ for five out of the 14 knots.  Such a high speed is comparable to the typical Si velocity seen in SNe Ia near maximum light.  Later, \cite{Millard:2020} measured precise Doppler shifts of Si He$\alpha$ lines for metal-rich ejecta knots, using high-resolution X-ray spectra with the Chandra HETG.  They found that some of the knots seem to be expanding nearly freely.  In addition, 8 out of the 15 ejecta knots show a statistically significant (at the 90\% confidence level) redshifted spectrum, but only two show blueshifted spectra.  This may suggest an asymmetry in the ejecta distribution in Kepler's SNR along the line of sight, although a larger sample size is required to confirm this interpretation.  Using the XMM-Newton RGS, \cite{Kasuga:2021} measured Doppler velocities of N and O lines which are dominated by the CSM.  The lines are overall blueshifted in a range of 0--500\,km\,s$^{-1}$.  On the other hand, the central bar structure shows an interesting spatial variation that the northwestern and southeastern halves are blueshifted and redshifted, respectively.  Such velocity structures are consistent with previous optical measurements \cite{Blair:1991}, implying a torus-like shape of the CSM distribution.  

{\it Tycho's SNR}: The remnant of SN~1572 --- a confirmed normal Type Ia SN, evidenced by the light-echo spectrum obtained with modern instruments \cite{Krause:2008}.   \cite{Sato:2017a} measured Doppler velocities for 27 compact X-ray knots using non-dispersive CCDs aboard Chandra.  The highest velocity knots are located near the center, while the low velocity ones appear near the edge as expected for a spherical expansion.  The typical velocities of the redshifted and blueshifted knots are $\sim$7800\,km\,s$^{-1}$ and $\sim$5000\,km\,s$^{-1}$, respectively.  Recently, \cite{Millard:2022} presented velocities of 59 metal-rich ejecta knots, based on Chandra HETG observations.  As a result, the distribution of space velocities throughout the remnant suggests that the southeast quadrant generally expands faster than the rest of the SNR.  Also, blueshifted knots are projected more in the northern shell, while redshifted knots are more in the southern shell, suggesting asymmetries in the CSM along the line-of-sight.

\section{Elemental Abundances of the CSM}
\label{sec:csm}

Core-collapse SNe and some of Type Ia SNe occur in a dense environment, i.e., CSM, created by mass-loss of massive progenitor stars.  After the SN explosion, the SN ejecta first interact with the surrounding CSM, giving rise to a variety of intense radiation.  This emission provides us with an excellent opportunity to study the nature of the progenitor star.  Below, we will briefly summarise results from high-resolution X-ray spectroscopy of the shock-heated CSM.

We can learn much about the CSM from light curves, energy spectrum (absorption, line profile, line intensity ratios, and their time variation), and polarization.  Of these, elemental abundances can be well determined by high-resolution spectroscopy, and have vital information to infer the evolutionary state of the progenitor star just before the SN explosion.  In particular, the relative abundances of C, N, and O are important, because the C/N/O ratio in the CSM changes significantly with the degree of the CNO processing that takes place in the H-rich envelope.  The relative C/N/O abundances of the stellar surface significantly vary with the evolutionary stage: the solar ratio (1/0.25/3 in number) is expected at the main sequence phase, and then a mild N over-abundance and O depletion are expected at a red supergiant (RSG) phase, and finally an extreme N over-abundance and O depletion (approaching 1/30/0.5 at the CNO equilibrium) are expected at a luminous blue variable or Wolf-Rayet (WR) phase \cite{Lamers:2001}.  Detection of the RSG-like abundance pattern suggests that the progenitor star was either a RSG star, or a LBV/WR star with a CSM ejected during a RSG phase; it is indeed not rare that debris of RGS winds are observed at a distance of a few pc around LBV/WR stars \cite{Smith:1997}.  On the other hand, detection of the CSM with the CNO-equilibrium abundance immediately tells us that the progenitor star was a LBV/WR star at the time of explosion. 

\begin{table}
\caption{Properties of CSM detected in SNRs}
\label{tab:CSM}       
\begin{tabular}{p{2.5cm}p{2cm}p{2cm}p{2cm}p{2cm}p{2cm}}
\hline\noalign{\smallskip}
Name & SN Type & N/O (solar) & $R_{\rm CSM}^a$ (pc) & $M_{\rm CSM}^{b}$ ($M_\odot$) & References \\
\noalign{\smallskip}\svhline\noalign{\smallskip}
SN~1987A & CC & 8 & 0.2 & 0.1 & \cite{Alp:2021,Sun:2021,McCray:2016} \\
SN~1978K & CC & 12 & $\lesssim$0.05 & 1 & \cite{Chiba:2020,Ryder:2016,Kuncarayakti:2016} \\
RX~J1713.7$-$3946 & CC & 7 & 4--9 & 0.002 & \cite{Tateishi:2021} \\
G296.1$-$0.5 & CC & 4 & 12 & 15 & \cite{Castro:2011,Tanaka:2022} \\
G292.0$+$1.8 & CC & 1 & 7 & 1.7 & \cite{Vink:2004,Bhalerao:2015} \\
Kepler's SNR & Ia & 1--6 & 3 & 0.3 & \cite{Blair:2007,Katsuda:2015,Kasuga:2021} \\
N103B & Ia & 0.5 & 3 & 3 & \cite{Williams:2014,Blair:2020,Yamaguchi:2021} \\
\noalign{\smallskip}\hline\noalign{\smallskip}
\end{tabular}
$^a$Distance between the CSM location and the explosion point.  $^b$Mass of the total CSM estimated in the entire remnant (SN~1978K, G296.1$-$0.5, Kepler's SNR, and N103B), or mass of peculiar CSM feature, i.e., equatorial rings for SN~1987A and G292.0$+$1.8, and the CSM knot (K1) for RX~J1713.7$-$3946.  
\end{table}

Evidence for X-ray emitting CSMs has been found in a number of SNRs.  Their observational properties are summarised in Table~\ref{tab:CSM}.  Most of N/O ratios of the CSM observed in core-collapse SNRs are mildly elevated from the solar value.  This is partly due to the observational bias, given that the identification of the CSM is mostly based on the N overabundance.  Therefore, there remains a possibility that we are missing CSMs with near solar abundances.  On the other hand, we can say that CSMs with the CNO-equilibrated abundance (extreme N enhancement) is rare in SNRs.  The N/O ratio can be used to infer the progenitor mass, because the N/O ratio is positively correlated with the progenitor mass.  For example, the N/O ratios for the CSM in SN~1978K and RX~J1713.7$-$3946 were measured to be $\sim$12 and $\sim$7 solar values, respectively, which led to estimates of their progenitor masses to be 10--20\,$M_\odot$ \cite{Chiba:2020} and 15--20\,$M_\odot$ \cite{Tateishi:2021}.  However, we should be cautious about this conclusion, because the N/O ratio depends not only on the stellar mass, but also on the stellar rotation and effects of binaries.  The best-studied CSM assiciated with SNRs would be the CSM ring in SN~1987A (the ER defined in section \ref{sec:global_ejecta}).  Recently, \cite{Sun:2021} reported an interesting finding that the metal abundances of N, O, Ne, and Mg significantly declined in the past few years.  This may be the result that the blast wave left the dense CSM ring and the relative contribution from the H II region compared with the dense CSM became larger than before.  The CSM in G292.0+1.8 is one of the rare class of CSM showing exceptionally low (near solar) N/O ratio.  It was identified as the CSM from its peculiar morphology, i.e., the equatorial belt of the SNR.  Its origin may be the stellar wind ejected before the CNO-processed material was dredged up to the surface of the progenitor star.  It may be also possible that a companion star is the main contributor to the equatorial belt, especially because the ring-like CSM feature such as the equatorial belt in G292.0+1.8 may have originated from binary interactions \cite{Morris:2009}.  

In case of Type Ia SNRs, the CSM detection itself is important, because it strongly suggests the presence of a companion star, favoring the single-degenerate scenario rather than the double-degenerate scenario for the progenitor system.  It is reasonable to assume that CSMs within Ia SNRs come from a donor star, based on a simple timescale estimates.  The CSM located at a distance of a few pc must have been blown some 10$^4$\,yr ago (a few pc divided by the CSM velocity of a few 100\,km\,s$^{-1}$), and this CSM should originate from a non-degenerate star (e.g., asymptotic giant branch or red giant star) judging from the elemental abundances.  On the other hand, it takes $\sim$10$^6$\,yr for a white dwarf to reach the Chandrasekhar mass by accretion from a donor star, so that the exploding star should have been a white dwarf when the CSM was ejected (some 10$^4$\,yr before the explosion), and thus the exploding star cannot blow the CSM observed.  In this way, the CSM associated with Type Ia SNRs allows us to study the donor star.

Kepler's SNR and N103B in the LMC are prototypical Ia SNRs with CSM.  The N/O ratios in the CSM were measured in both optical and X-ray wavelengths.  Given that the N/O ratio on the surface of stars with masses below 8\,M$_\odot$ generally increases with increasing stellar mass \cite{Karakas:2016}, we can constrain the mass of the donor star from the N/O ratio.  The CSM observed in Kepler's SNR exhibits an over-abundance of N, which suggests a relatively massive donor star, possibly an ABG star.  In this case, a surviving companion star is expected to remain very bright ($\sim$10$^3 L_\odot$) for at least 10$^5$\,yr after the SN explosion \cite{Marietta:2000}.  However, no such bright surviving companion star has been found near the center of the remnant, posing a challenge to the single-degenerate scenario.  On the other hand, \cite{Blair:2007,Blair:2020,Dopita:2019} have shown that the N enhancements in the CSM knots of Kepler's SNR are actually consistent with those expected in the local ISM at its location only $\sim$3\,kpc away from the Galactic center.  Moreover, a recent RGS observation \cite{Kasuga:2021} estimated the N abundance of the CSM to be only 1--2 solar, which conflicts with past X-ray (and optical) results, providing us with yet another possibility that the metal abundance of the CSM is even lower than the local ISM.  If true, a massive AGB donor star may not be required, which would suggest a different progenitor system.  In this context, it is important to revisit the CSM abundance in Kepler's SNR with the upcoming XRISM \cite{Tashiro:2020}.  Lastly, the low N/O ratio measured in N103B is (also) consistent with the local ISM of the LMC \cite{Blair:2020,Yamaguchi:2021}.  \cite{Williams:2014} argued that the CSM originated from a mass loss prior to the first dredge-up of material in a donor star.


\section{Probing Interstellar Dust and Solar Planets' Atmospheres by Extinction of X-Ray Emission from the Crab Nebula}
\label{sec:extinction}

Bright SNRs provide us with unique opportunities to unveil the physical and chemical properties of the diffuse ISM through measurements of absorption lines and edges in the X-ray spectrum.  The Crab Nebula, the remnant of SN~1054, would be the best object for this study, because it is one of the brightest sources in the whole X-ray sky with its flux being relatively steady, and it has a simple power-law spectrum.  In addition, the angular size of the Crab Nebula is small enough for the XMM-Newton RGS to take a spectrum with good spectral resolution.  

\begin{figure}[htb]
\includegraphics[scale=0.4]{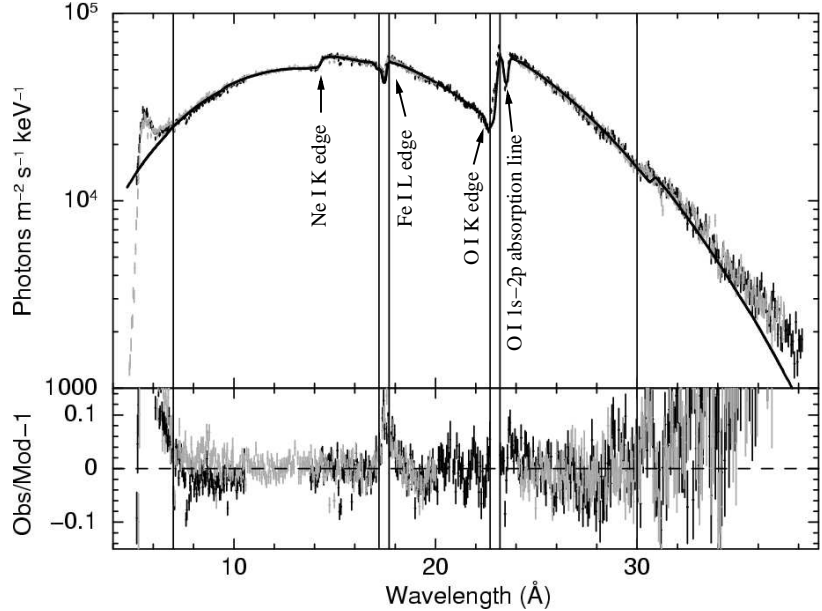}
\caption{RGS spectrum from the Crab Nebula.  RGS1 and RGS2 spectra are separately shown in dark and light colors, respectively.  Two narrow regions near the O K-edge and Fe L-edge (indicated by vertical lines) are excluded in the fitting.  The best-fit model, which consists of an absorbed (pure gas), curvature-corrected power-law component, is shown as a solid line.  Lower panel shows the residuals.  This figure is taken from \cite{Kaastra:2009}.  Reproduced with permission \textcircled{C} ESO.}
\label{fig:Crab_rgs}       
\end{figure}

\cite{Kaastra:2009} performed high-resolution X-ray spectroscopy for the Crab Nebula using the XMM-Newton RGS, following earlier works with Einstein \cite{Schattenburg:1980,Schattenburg:1986} and Chandra \cite{Weisskopf:2004}.  As shown in Fig.~\ref{fig:Crab_rgs}, the RGS successfully detected strong $1s-2p$ absorption lines of neutral O, absorption K-shell edges of neutral O and Ne, and L-shell edges of neutral Fe.  These absorption features are due to the ISM, allowing one to measure column densities and abundances relative to H of N, O, Ne, Mg, and Fe in the ISM.  It was found that N/H and O/H abundances fully agree with the solar values \cite{Lodders:2003}, and Mg/H and Fe/H abundances are slightly lower than the solar values.  On the other hand, the Ne/H abundance is elevated by a factor of 1.7, which is equivalent to the Ne/O number ratio of 0.26.  Interestingly, this Ne/O ratio is in excellent agreement with those obtained in the Orion Nebula and of B-stars in that nebula.  In fact, the Ne abundance of the Sun has been subject to a relatively large uncertainty, because Ne lacks detectable photospheric lines in cool stars like the Sun.  A more recent standard solar abundance \cite{Lodders:2009} provides the Ne/O abundance ratio of 0.21, which is still smaller than, but closer to the value obtained for the ISM toward the Crab Nebula.

In addition, edge spectral regions can be a unique tool for probing different phases of the ISM, i.e., gas and solid (dust) phases, because extinction cross sections are different between gas and dust at edge regions.  In particular, X-ray absorption fine structures (XAFSs) are unique fingerprints of dust (see, e.g., \cite{Psaradaki:2020} for the study of O K-edge structures).  From simultaneous modeling of absorption edges of O K, Fe L, and Ne K, \cite{Kaastra:2009} suggested that the dust fraction (dust-to-gas mass ratio) in the line of sight toward the Crab Nebula is $\sim$70\% of the typical value in the local ISM \cite{Wilms:2000}.  The relatively low abundance of dust was also inferred from the dust-scattering optical depth measured by Chandra imaging of the dust scattering halo around the Crab Nebula \cite{Seward:2006}.  In addition to the edge features, the line centroid of the O absorption line was measured at 23.47\,\AA\, with the RGS \cite{Kaastra:2009}.  This is in between the atomic O I line at 23.51\,\AA\, and hematite (Fe$_2$O$_3$) line at 23.43\,\AA, suggesting approximately equal contributions from atoms and hematite of a ferric compound.  Such a mixture is in rough agreement with the $\sim$30\% of O is bound in dust as derived from the wide-band RGS spectral modeling.  

It is interesting to note that the location of the O K edge was measured to be 541\,eV (22.917\,\AA), which is shifted blueward by 3\,eV or 0.128\,\AA\, from 538\,eV (23.045\,\AA) expected for the neutral O K-edge.  In addition to the O K-edge, Ne I K and Fe L-edges also appear to be blueshifted to some extent, which was suggested to be caused by neglect of possible contaminations by highly ionised O.  It is not easy to specify the cause(s) of the K-edge shifts, because the O K-edge structure is complex due to many absorption features by neutrals, ionised gas, and dust, and also due to the fact that the effective spectral resolution is not very high because of the spatial extent of the Crab Nebula.  Nonetheless, we point out one possibly important (missing) effect that the RGS analysis by \cite{Kaastra:2009} ignored scattering contributions to the dust extinction.  Although this treatment is a usual assumption in the X-ray astronomy \cite{Wilms:2000}, it is actually not a good approximation; \cite{Hoffman:2016} claimed that the scattering cross section by dust grains exceeds that of photoelectric absorption above 1\,keV.  The effect of dust scattering could significantly change the edge structures.  Whatever the origin, we will be able to better understand the shifts of edges with the XRISM Resolve which will provide us with edge and absorption structures in unprecedented detail.

Later, \cite{Weisskopf:2004,Weisskopf:2011} analysed the Chandra LETG spectrum of the Crab Nebula.  The superb angular resolution of Chandra enabled distinguishing the Crab pulsar from the surrounding nebula along the cross-dispersion direction, making the spectral analysis simple and robust.  The line spacing of the LETG is however as large as 10$^{4}$\,\AA, which is 25 times larger than that of the RGS, resulting in substantial spectral blurring due to the spatial extent of the Crab Nebula along the dispersion axis.  Nonetheless, the Chandra LETG successfully detected the O K-edge.  The O/H abundance of the ISM toward the Crab Nebula was estimated to be (5.28$\pm$0.28)$\times$10$^{-4}$, which is consistent with the RGS measurement \cite{Kaastra:2009}.

For the Chandra LETG spectrum, \cite{Weisskopf:2004,Weisskopf:2011} considered the dust scattering effect, by simply multiplying an exponential-decay term, exp[$-\tau_{\rm sca} (E)$], to their spectral-fitting model, where the energy dependence of $\tau_{\rm sca} (E)$ was assumed to be $E^{-2}$ which is the so-called Rayleigh-Gans approximation (note however that this assumption is known to be too simple at the energy range of the Chandra LETG).  The scattering component was not significantly required initially \cite{Weisskopf:2004}, but it turned out to be required at almost 4-$\sigma$ level by the long-exposure data combined with the most recent response files \cite{Weisskopf:2011}.  The best-fit scattering optical depth was obtained to be $\tau_{\rm sca} = 0.147\pm0.043$ at 1\,keV.  

On the other hand, the dust-scattering optical depth has been also measured by the fractional halo intensity as $\tau_{\rm sca}(E) = {\rm ln}[1 + N_{\rm halo}(E)/N_{\rm src}(E)]$ at the photon energy $E$, where $N_{\rm halo}$ and $N_{\rm src}$ represent fluxes of the dust-scattering halo and the central source, respectively.  Thus-derived optical depths toward the Crab Nebula vary from observation to observation: $\tau_{\rm sca} \sim 0.05$ in 0.4--2.1\,keV from Chandra \cite{Seward:2006}, $\tau_{\rm sca} \sim 0.14$ at 1\,keV from Einstein \cite{Mauche:1989}, $\tau_{\rm sca} \sim 0.09$ at 1\,keV from ROSAT \cite{Predehl:1995}.  The optical depth obtained from the LETG spectroscopy as shown in the previous paragraph agrees with the result from the Einstein imaging, but are smaller than the other results.  Systematic observations of dust scattering halos around Galactic bright point sources revealed strong positive correlations for $\tau_{\rm sca}$--$N_{\rm H}$ and $\tau_{\rm sca}$--$A_{\rm V}$: $\tau_{\rm sca}(1~{\rm keV}) = 0.025 \times (N_{\rm H}/10^{21} {\rm cm}^{-2})$ and $\tau_{\rm sca}(1~{\rm keV}) = 0.041 \times (A_{\rm V}/1~{\rm mag})$ obtained for the ``MRN" dust size distribution model \cite{Valencic:2015}.  If we adopt the canonical values of $N_{\rm H} = 3.2 \times 10^{21}$\,cm$^{-2}$ and $A_{\rm V} = 1.6$\,mag for the Crab Nebula, the expected values of $\tau_{\rm sca}$ are obtained to be 0.08 and 0.066 from $\tau_{\rm sca}$--$N_{\rm H}$ and $\tau_{\rm sca}$--$A_{\rm V}$ relations, respectively.  These values are expected for a typical dust depletion in the ISM.  The smaller-than-expected values from the Chandra imaging and the RGS spectroscopy suggest less dust to the Crab Nebula, and the larger-than-expected values from other measurements suggest more dust than the typical ISM.

The dust depletion factor is also sensitive to the relative depth of the edge to the absorption line \cite{Pinto:2013}.  Dust is usually responsible for a deep edge, whereas the gas provides strong lines as well as the edge, so that a deep edge together with a shallow absorption line indicates a large dust fraction.  As can be seen in Fig.~\ref{fig:Crab_rgs}, the RGS spectrum of the Crab Nebula shows a much deeper O K-edge than the absorption line \cite{Kaastra:2009}.  This edge-to-line depth ratio is indeed the highest among the sources exhibited in \cite{Pinto:2013}, implying that the dust depletion factor toward the Crab Nebula is the highest among those measured by \cite{Pinto:2013}, meaning that more than 20\% of O is depleted into dust.  This is consistent with the result from the RGS spectroscopy \cite{Kaastra:2009}.  However, at this moment, there remain some uncertainties on the RGS spectrum of the Crab Nebula, as described above.  Therefore, further refinements of the dust scattering optical depth ($\tau_{\rm sca}$) are eagerly awaited for future observations.

Another intriguing application of the occultation technique with the Crab Nebula and other very bright point sources is to measure vertical density profiles of the Earth's atmosphere.  X-ray astronomy satellites, which point at a selected target during each observation in low-Earth orbits ($\sim$600-km height above the sea level), observe target's setting and rising behind the Earth every orbit.  During the setting (and rising), X-rays from the target experience substantial extinction by the atmosphere.  Therefore, we see that the X-ray intensity gradually decreases (increases) as the line-of-sight approaches (departs from) the Earth's surface.  By modeling the atmospheric attenuation of X-ray photons during each occultation (either spectral fitting or light curve fitting), we can measure vertical density profiles of the atmosphere.  So far, this technique has not allowed one to measure the composition of the atmosphere, but enabled measuring the total atomic (sum of atoms and molecules) number densities along the line of sight.  

A pioneering work on X-ray occultation sounding of the upper atmosphere of the Earth was done by \cite{Determan:2007} who demonstrated the analysis of atmospheric occultations of the Crab Nebula and Cygnus X-2 observed with the RXTE satellite.  Renewed interest in this work was recently stimulated with the analyses of the archival data of the Crab Nebula with Suzaku and Hitomi \cite{Katsuda:2021}, which was followed by several subsequent studies \cite{Yu:2022a,Yu:2022b,Xue:2023,Katsuda:2023}.  Systematic measurements of the vertical density profiles of the upper atmosphere with the Insight-HXMT satellite \cite{Xue:2023} showed good agreements with the latest version of the major atmospheric model, i.e., NRLMSIS 2.0 \cite{Emmert:2021}.  Also, \cite{Katsuda:2023} recently showed that the density of the upper atmosphere has been gradually decreasing from 1994 to 2022.  This was interpreted as the result of ``greenhouse cooling" in the upper atmosphere due to the increasing greenhouse gases.  

Atmospheres of other solar planets can be also investigated with the X-ray occultation technique.  The atmospheric thickness of Titan, i.e., the Saturn's largest satellite, was measured from the transit of the Crab Nebula on January 05, 2003 \cite{Mori:2004}.  An atmospheric density profile at Mars was measured by occultations of $\sim$10\,keV X-rays from Scorpius X-1, using the SEP instrument on the MAVEN spacecraft \cite{Rahmati:2020}.

The future looks bright for this study.  It is of great importance to keep monitoring the density in the upper atmosphere to predict climate changes and to refine orbital decays and/or lifetimes of the satellites.  A great advantage with this technique is that X-ray astronomy satellites in low Earth orbits routinely (can not avoid to) observe the settings and risings of the targets, providing us with rich data sets to measure atmospheric densities.  Moreover, as demonstrated in \cite{Katsuda:2021}, the unprecedented spectral resolution of $\sim$5\,eV in 0.3--12 keV of the XRISM Resolve will allow us to measure not only total densities but also compositions from absorption edges of N, O, and possibly Ar, just like the composition measurements from the extinction by the ISM.  In particular, the O/N$_2$ ratio in the upper atmosphere will be an interesting parameter, as this ratio is thought to influence the ionospheric electron density and is expected to probe the thermosphere-tide coupling process.  It will be also interesting to study short-term variability of the upper atmosphere that could be caused by both the solar activity and dynamics of the lower atmosphere.


%
%
%

\begin{acknowledgement}
I am grateful to Dr.\ Casey T.\ DeRoo and Dr.\ Takashi Okajima for their kind explanations on X-ray grating spectrometers and X-ray optics.  I would also like to thank Dr.\ Hiroyuki Uchida for a long-term collaboration on high-resolution X-ray spectroscopy of SNRs and for providing me with the idea for Figure~\ref{fig:papers_cum}.  This work was supported by the Grants-in-Aid for Scientific Research from the Japanese Ministry of Education, Culture, Sports, Science, and Technology (MEXT) of Japan, No. JP20K20935, JP20H00174, and JP21H01121.  I also deeply appreciate the Observational Astrophysics Institute at Saitama University for supporting the research fund.  

\end{acknowledgement}






\end{document}